\documentclass[twocolumn,showpacs,preprintnumbers,amsmath,amssymb]{revtex4}

\usepackage{graphicx}
\usepackage{dcolumn}
\usepackage{bm}

\begin{document}

\title{Probing of a parametrically pumped magnon gas\\ with a non-resonant packet of traveling spin waves}

\author{T. Neumann}
 \email{neumannt@physik.uni-kl.de}
\author{A. A. Serga}
\author{B. Hillebrands}
\affiliation{
Fachbereich Physik and Forschungszentrum OPTIMAS\\
Technische Universit\"at Kaiserslautern, 67663 Kaiserslautern, Germany}

\date{\today}

\begin{abstract}
The magnon gas created by spatially localized longitudinal parametric pumping in an yttrium-iron-garnet
film is probed by a traveling packet of spin waves non-resonant with the pumping field. The analysis of
the influence of the magnon gas on the amplitude and phase of the propagating spin waves allows to
determine characteristic properties of the parametrically pumped magnon gas. A simple theoretical model
is proposed from which the magnon density in the pumping region is calculated.
\end{abstract}


\maketitle

Parallel parametric pumping of spin waves is a widely used and established technique to control the
magnon density in different areas of the spin-wave spectrum. It is used both in experiments on
fundamental properties of magnetic excitations and for the amplification and processing of microwave
signals \cite{Pat06}. It has been used to realize spin-wave wavefront reversal \cite{Mel04, Mel05}, to
amplify spin-wave solitons and bullets \cite{Ser05}, to generate M\"{o}bius solitons \cite{Dem03} as
well as to create Bose-Einstein condensates of magnons at room temperature \cite{Dem06}. The long-time
storage and recovery of microwave pulses \cite{Ser07-a}, the microwave spectral analysis \cite{Sch08},
and the shaping of microwave pulses \cite{Ser07} has been achieved with this technique.

Parametric pumping amplifies many different spin-wave groups simultaneously. In the process the density
of spin waves is increased initially at half the pumping frequency $f_{\rm p} / 2$. These amplified spin
waves form an overheated region of the magnon gas which then thermalizes spreading its energy over the
whole spin-wave spectrum \cite{Dem07}. For long enough pumping a quasi-stationary magnon gas state
obeying a Bose-Einstein distribution with a concentration of magnons near the bottom of the spin-wave
spectrum is created.

Due to the high overall density of the magnon gas spatially localized in the pumping area the static
magnetization there is decreased and a magnetic inhomogeneity is formed. From this {\it magnon barrier}
\cite{Ser07} propagating spin waves can scatter similar to the scattering of spin waves on a dc-current
induced local magnetic inhomogeneity \cite{Kos07,Dem08}. Also, direct scattering of traveling spin waves
on the parametrically excited magnons can occure \cite{Mel05-a}. As a consequence, the experiments
performed so far on the interaction of the traveling spin waves with parametric pumping had to take into
account two competing effects: while the pumping field tends to increase the spin-wave amplitude the
magnon barrier decreases it via scattering.

In the current paper, we separate these effects by using {\it non-resonant} spin-wave pulses with a
frequency $f_{\rm s} \neq f_{\rm p} / 2$. We study the effects of the magnon barrier on the amplitude of
a traveling spin-wave independent of any amplification caused by direct interaction with the applied
external pumping field.

\begin{figure}[t]
\includegraphics[height = 39ex]{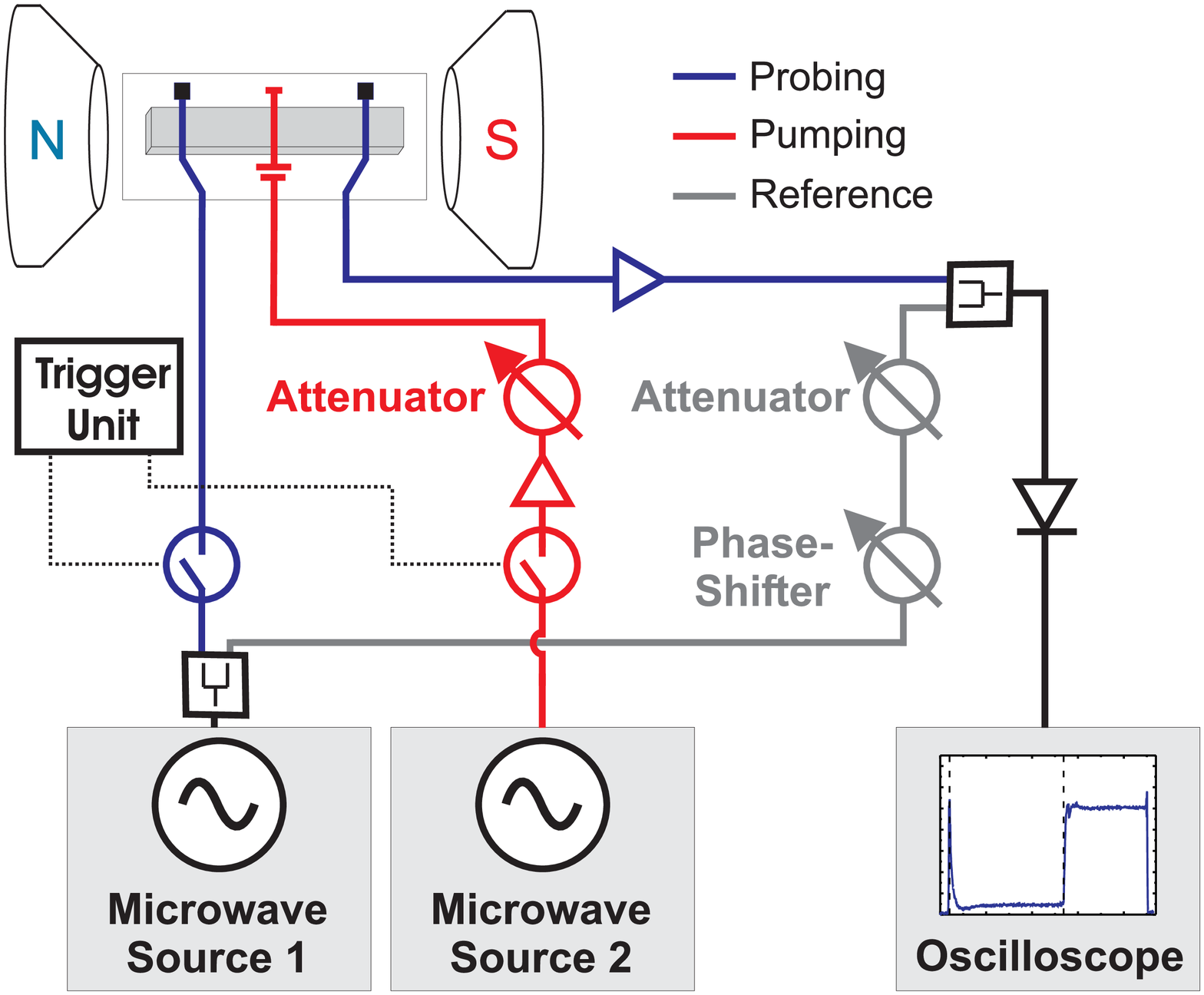}
\caption{\label{fig:aufbau} (Color online) Sketch of the experimental setup.}
\end{figure}

The magnon barrier also influences the spin-wave phase by introducing an additional phase shift. It is
known, that this phase shift is of crucial importance for the amplification by non-adiabatic localized
parametric pumping \cite{Dem03,Ser07}. Our measurements have shown that this additionally accumulated
phase is more sensitive to the barrier characteristics than the amplitude at intermediate to high
pumping powers. Since these regimes are interesting both for practice and fundamental studies (e..g for
the generation of Bose-Einstein condensates), a major part of the current paper is devoted to the
investigation of the spin-wave phase.

By measuring the pure {\it scattering} of the spin-wave amplitude on the magnon barrier as well as the
{\it phase accumulation} of the spin wave upon propagating through the pumping region physically
relevant characteristic properties of the magnon barrier can be obtained. We illustrate the potential of
probing a parametrically generated magnon gas with non-resonant propagating spin-wave packets by
estimating the magnon density in the pumping area based on a simple model.

A schematic view of the experimental setup is shown in Fig. \ref{fig:aufbau}. The microwave source "1"
generates a cw-signal at $f_{\rm s} = 6.82~{\rm GHz}$. A triggered microwave switch transforms this
signal into a $8~\mu{\rm s}$ long microwave pulse which is sent to an input transducer placed across a
$7.8~\mu{\rm m}$ thick and $1.5~{\rm mm}$ wide yttrium iron garnet (YIG) sample. Due to the chosen
geometry the microwave signal excites a packet of so called backward volume spin waves which propagate
with a wave vector ${\rm k} = 150~{\rm cm}^{-1}$ parallel to the applied bias magnetic field $\rm{H}=
1750~{\rm Oe}$. An output antenna $8~{\rm mm}$ from the input antenna picks up the spin-wave pulse after
it has passed through the film.

\begin{figure}[t]
\includegraphics[height = 21ex]{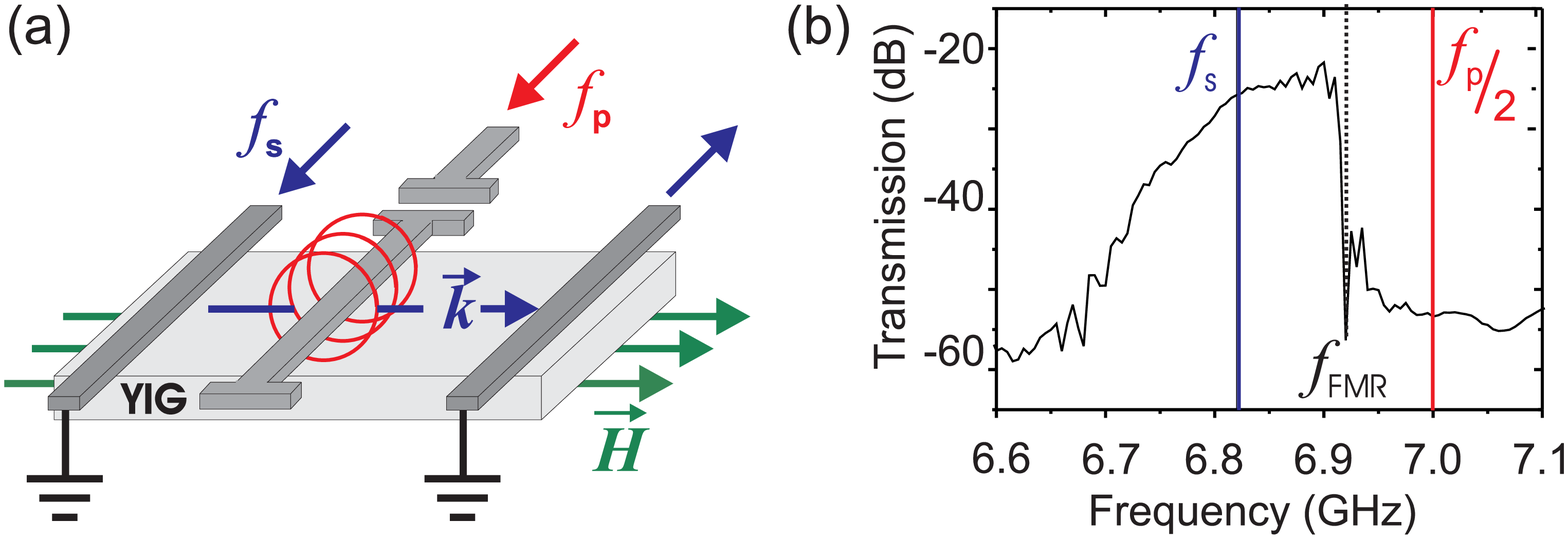}
\caption{\label{fig:section} (Color online) (a) Experimental section. (b) Microwave cw-transmission
characteristics.}
\end{figure}

A $50~\mu{\rm m}$-wide microstrip resonator half-way between the two microstrip transducers (see fig.
\ref{fig:section}(a)) allows effective, local parametric pumping. The microwave source "2" drives the
resonator with a $5~\mu{\rm s}$ long pumping pulse at a frequency $f_{\rm p}=14.00~{\rm GHz}$. Note,
that there is a discrepancy of $\delta f = f_{\rm p} / 2 - f_{\rm s} = 180~{\rm MHz}$ by which the
resonance condition for direct amplification of the propagating spin waves is {\it not} fulfilled. The
applied pumping power $P$ can be varied up to $80~{\rm W}$. In order to avoid overheating the sample at
such high powers a repetition rate of $2~{\rm ms}$ is chosen.

The transmission characteristics presented in Fig. \ref{fig:section}(b) show that the frequency of the
pumping pulse is above the frequency of ferromagnetic resonance $f_{\rm FMR}$ which makes sure that,
first, the pumping threshold is low \cite{Mel96} and, second, no propagating backward volume spin waves
are amplified. Simultaneously, the carrier frequency of the traveling spin-wave packet is slightly below
$f_{\rm FMR}$, which guarantees its good excitation and detection via the microstrip antennas.

The signal received at the output antenna can interfere with a phase-locked reference signal before
detection and observation in order to reconstruct its phase \cite{Ser06}.

\begin{figure}[t]
\includegraphics[height = 45ex]{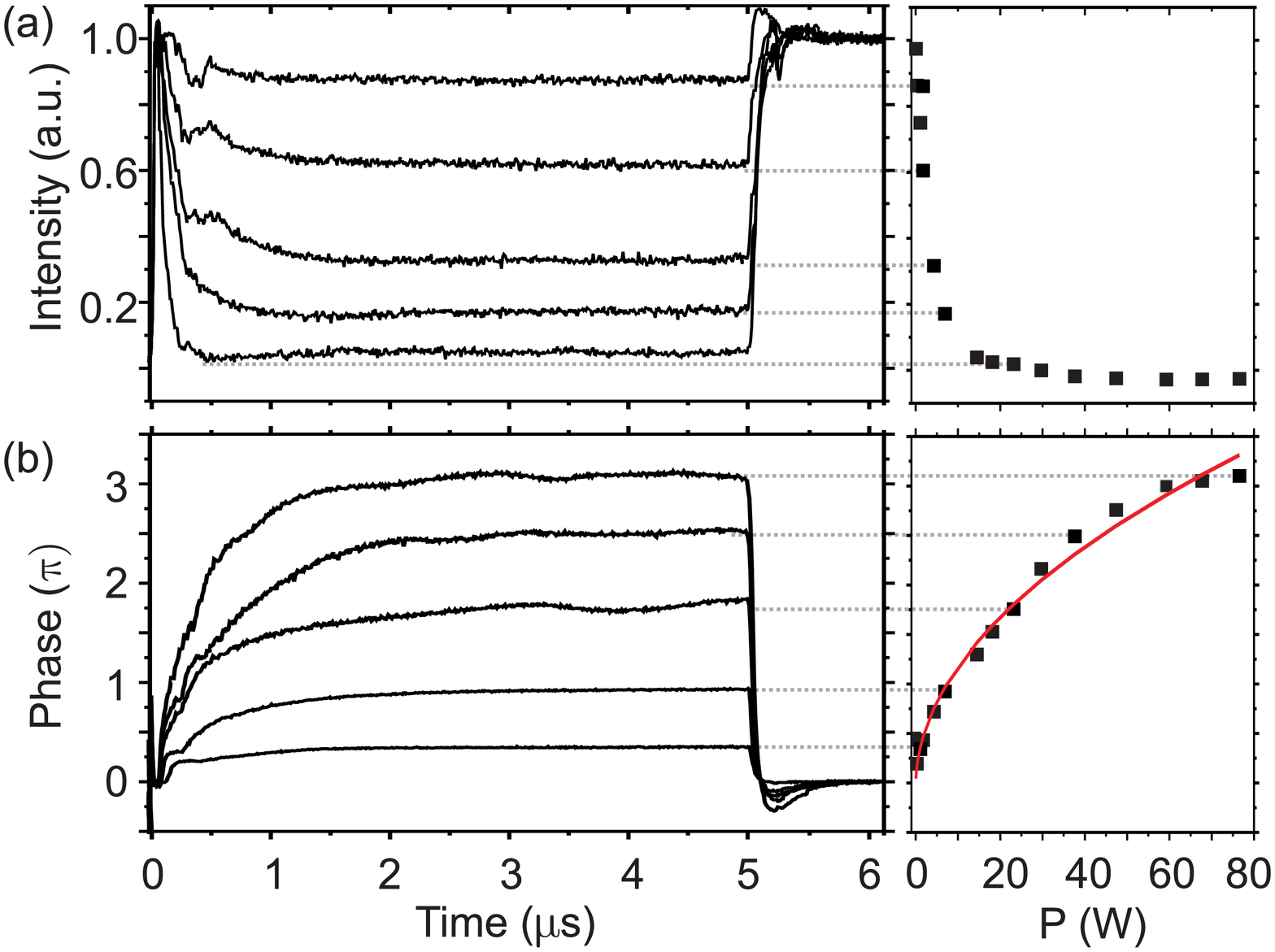}
\caption{\label{fig:results} (Color online) (a) Transmitted signal shapes for different pumping powers.
(b) Phase profiles for different pumping powers. The solid line in the right panel indicates the
corresponding calculated dependence. }
\end{figure}

The measured pulse shapes for different applied pumping powers are shown in the left panel of Fig.
\ref{fig:results}(a). We note three things: First of all, no signal amplification takes place since the
propagating spin wave is non-resonant with the external pumping. Secondly, the spin-wave signal is
suppressed when the pumping pulse is applied. Thirdly, the suppression increases monotonically with
increasing pumping power. In particular, already at intermediate pumping powers of $20~{\rm W}$ almost
complete signal suppression is achieved. This suppression does not happen instantaneously when the
pumping pulse is switched on, but takes up to $2~\mu{\rm s}$ to reach a stationary regime.

The observed suppression is understood as the result of scattering of the magnon barrier. With
increasing pumping power, the scattering increases together with the magnon barrier. As the magnon
barrier is created in the process of the thermalization of magnons excited at half the pumping frequency
a relatively slow passage into the stationary regime is expected.

The left panel of Fig. \ref{fig:results}(b) shows the measured phase profiles of the output pulse
relative to the reference pulse. As for the transmitted intensity, the accumulated phase slowly develops
towards a stationary value. This stationary phase depends monotonically on the applied pumping power.
However, if we compare the corresponding curves we note that the phase does not seem to saturate even at
high powers.

A straightforward model describes the observed phase of the propagating spin-wave in the stationary
regime.

In the process of pumping the magnon density in the area around the pumping resonator is increased. This
leads to a small reduction $\delta M$ of the magnetization $M$ from its initial value $M_{\rm s}=
139~{\rm G}$. This changes the wave vector of the propagating spin wave. The difference between the
undisturbed and disturbed wave vector integrated over the distance the spin wave propagates through the
pumping region gives the accumulated phase observed in the experiment.

Assume for simplicity that the parametrically generated magnons are homogeneously distributed in a
region of length $l$ around the central pumping resonator. Only in this region the magnetisation is
reduced and a phase $\phi$ is accumulated. As a consequence

\begin{equation}\label{Equ-phi}
\phi = \int \bigg( k(M_{\rm s})-k(M) \bigg) dx = \bigg(k(M_{\rm s})- k(M)\bigg) \cdot l.
\end{equation}

We linearize the dependence of the wave vector on the magnetization of the sample $k(M) = k(M_{\rm s}) +
c_1\cdot \delta M$ where the constant $c_1 = 18.3~{\rm cm}^{-1} {\rm G}^{-1}$ can be calculated from the
dispersion law and the experimental conditions. Every single parametrically generated magnon decreases
the magnetization by one Bohr magneton $\mu_{\rm B}$. As a result, if $N$ denotes the number of magnons
created by the pumping source and $V=l\cdot w\cdot d$ is the volume of the pumping region:
\begin{equation} \label{Equ-MagDens}
\phi = c_1 \cdot \mu_{\rm B} \cdot l \cdot \frac{N}{V} = \frac{c_1 \cdot \mu_{\rm B}}{w \cdot d} \cdot N
.
\end{equation}

The fact, that a stationary amplitude is observed, shows, that the magnon density does not grow
infinitely. It is limited by several mechanisms, primarily the phase mechanism \cite{Zak74}, which
reduces the effective coupling of the pumping field with the parametric magnons.

According to \cite{Zak74}, the stationary magnon number for a given wave vector taking into account is
given by
\begin{equation} \label{Equ-MagNumber}
N = \frac{1}{S} \sqrt{(h_{\rm p} V_{\rm coupl})^2 - \omega_r^2}
\end{equation}
where $S = \Big(\omega_{\rm M}/\omega_{\rm p} \Big)^2 \Big( \sqrt{\omega_{\rm p}^2 + \omega_{\rm M}^2} -
\omega_{\rm M}\Big)$ / 4, $\omega_{\rm p} = 4\pi f_{\rm p}$, $\omega_{\rm M} = \gamma \; 4\pi M_{\rm
s}$. $h_{\rm p}$ is the pumping magnetic field in the direction of the bias magnetic field, $V_{\rm
coupl}$ is the coupling coefficient, $\omega_{\rm r}$ is the relaxation frequency and $\gamma$ is the
gyromagnetic ratio. Since $h_{\rm p} \sim \sqrt{P}$ we get as final result
\begin{equation} \label{Equ-Final}
\phi = \frac{c_1\cdot \mu_{\rm B} \cdot l}{V\cdot S} \sqrt{c_2 P - \omega_{\rm r}^2} \approx \frac{c_1\cdot \mu_{\rm B} \cdot l}{V\cdot S} \sqrt{c_2 P}\\
\end{equation}
with $c_2$ a constant depending on the geometrical properties of the resonator and the coupling of the
spin waves to the external pumping field. The last approximation is justified for pumping powers not too
close to the pumping threshold which as it is the case in our experiments.

As Fig. \ref{fig:results}(b) shows, the presented model with the single free parameter $c_2$ fits the
experimental data well.

From the experimentally observed accumulated phase we can calculate
\begin{displaymath}
N = 5.5 \cdot 10^{13} \cdot \phi, \quad\quad \frac{N}{V} = 2.3 \cdot 10^{19}~{\rm cm^{-3}} \cdot \phi
\end{displaymath}
where we have estimated $l=0.2~{\rm mm}$ \cite{Mel99} to obtain the density of parametrically pumped
magnons. We note, that the later value agrees well with expected values for the magnon density of
$10^{18}-10^{19}~{\rm cm}^{-3}$ \cite{Mel05-a, Mel96}.

In conclusion, we have measured the transmission and phase accumulation of non-resonant spin waves
propagating through a parametrically generated magnon barrier. From the accumulated phase the absolute
number of magnons constituting the barrier was deduced and the total magnon density was estimated.

Overall, the method of probing the total density of a parametrically controlled magnon gas by the phase
of a non-resonant traveling spin wave possesses great potential due to its time-resolution and high
sensitivity. It is applicable to thin magnetic materials, e.g. thin magnetic films, where techniques
based on other physical effects fail.

Financial support by the Matcor Graduate School of Excellence and the DFG within the SFB/TRR 49 is
gratefully acknowledged.


\begin{thebibliography}{1}

\bibitem{Pat06} C. Patton, M. Wu, K.~R. Smith, V.~I. Vasyuchka, Ferroelectrics {\bf 342}, 101 (2006).

\bibitem{Mel04} G.~A. Melkov, V.~I. Vasyuchka, and Yu.~V. Kobljanskyj, Phys. Rev. B. {\bf 70}, 224407 (2004).

\bibitem{Mel05} G.~A. Melkov, V.~I. Vasyuchka, and A.~V. Chumak, J. Appl. Phys. {\bf 98}, 074908 (2005).

\bibitem{Ser05} A.~A. Serga, B. Hillebrands, S.~O. Demokritov, A.~N. Slavin, P. Wierzbicki, V.~I. Vasyuchka, O. Dzyapko, and A.~V. Chumak, Phys. Rev. Lett. {\bf 94}, 167202 (2005).

\bibitem{Dem03} S.~O. Demokritov, A.~A. Serga, V.~E. Demidov, B. Hillebrands, M.~P. Kostylev, and B.~A. Kalinikos, Nature {\bf 426}, 159 (2003).

\bibitem{Dem06} S.~O. Demokritov,  V.~E. Demidov, , G.~A. Melkov, A.~A. Serga, B. Hillebrands, and A.~N. Slavin, Nature {\bf 443}, 430 (2006).

\bibitem{Ser07-a} A.~A. Serga, A.~V.Chumak, A. Andr\'e, G.~A. Melkov, A.~N. Slavin, S.~O. Demokritov, and B. Hillebrands, Phys. Rev. Lett. {\bf 99}, 227202 (2007).

\bibitem{Sch08} S. Sch\"afer, A.~V. Chumak, A.~A. Serga, and B. Hillebrands, Appl. Phys. Lett. {\bf 92}, 162514 (2008).

\bibitem{Ser07} A.~A. Serga, T. Schneider, B. Hillebrands, M.~P. Kostylev, and A.~N. Slavin, Appl. Phys. Lett. {\bf 90}, 022502 (2007).

\bibitem{Dem07} V.~E. Demidov, O. Dzyapko, S.~O. Demokritov, G.~A.  Melkov, and A.~N. Slavin, Phys. Rev. Lett. {\bf 99}, 037205 (2007).

\bibitem{Kos07} M.~P. Kostylev, A.~A. Serga, T. Schneider, T. Neumann, B. Leven, B. Hillebrands, and R.~L. Stamps, Phys. Rev. B {\bf 76}, 184419 (2007).

\bibitem{Dem08} V.~E. Demidov, U.~H. Hansen, and S.~O. Demokritov, Phys. Rev. B {\bf 78}, 054410 (2008).

\bibitem{Mel05-a} G.~A. Melkov, Yu.~V. Kobljanskyj, and O. Dzyapko, Ukr .Fiz. Zhurn. {\bf 50}, A5 (2005).

\bibitem{Ser06} A.~A. Serga, T. Schneider, B. Hillebrands, S.~O. Demokritov, and M.~P. Kostylev, Appl. Phys. Lett. {\bf 89}, 063506 (2006).

\bibitem{Zak74} V.~E. Zakharov, V.~S. L'vov, and S.~S. Starobinets, Sov. Phys. Usp. {\bf 17}, 6 (1974).

\bibitem{Mel99} G.~A. Melkov, A.~A. Serga, A.~N. Slavin, V.~S. Tiberkevich, A.~N. Oleinik, and A.~V. Bagada, JETP {\bf 89}, 1189 (1999).

\bibitem{Mel96} A.~G. Gurevich and G.~A. Melkov (CRC Press, Inc., 1996).

\end{thebibliography}
\end{document}